\documentstyle[11pt]{article}
\begin{document}
\vspace{.3in}
\title{
%StartAbstract
Multidimensional Pattern Formation Has an Infinite Number 
of Constants of Motion
%StopAbstract
}
\author{
%StartAbstract
Mark B. Mineev--Weinstein
%StopAbstract
\\[.3in]
Courant Institute of Mathematical Sciences\\ 251 Mercer Street\\ 
New York University\\ New York, NY 10012\\and \\
Center for Nonlinear Studies, MS--B258\\Los Alamos National Laboratory\\
Los Alamos, NM 87545 }
\maketitle
\begin{abstract}
%StartAbstract
Extending our previous work on 2D growth for the Laplace equation we
study here {\it multidimensional} growth for {\it arbitrary elliptic}
equations, describing inhomogeneous and anisotropic pattern formations
processes.  We find that these nonlinear processes are governed by an
infinite number of conservation laws. Moreover, in many cases {\it all
dynamics of the interface can be reduced to the linear time--dependence
of only one ``moment" $M_0$} which corresponds to the changing volume
while {\it all higher moments, $M_l$, are constant in time. These
moments have a purely geometrical nature}, and thus carry information
about the moving shape.  These conserved quantities (eqs.~(7) and (8)
of this article) are interpreted as coefficients of the multipole
expansion of the Newtonian potential created by the mass uniformly
occupying the domain enclosing the moving interface.  Thus the question
of how to recover the moving shape using these conserved quantities is
reduced to the classical inverse potential problem of reconstructing
the shape of a body from its exterior gravitational potential.  Our
results also suggest the possibility of controlling a moving interface
by appropriate varying the location and strength of sources and sinks.
%StopAbstract
\end{abstract}
\vspace{.3in}
\hspace{.4in}
\pagebreak

Many seemingly different pattern formation processes have much in common, both 
in their mathematical description and in their physical behavior. Among them 
are the famous Stefan problem (freezing of liquid), flow through porous media,
the Rayleigh--Taylor instability, electrodeposition of metals, corrosion, 
combustion, growth of bacterial colonies, dynamics of earth cracks, 
diffusion--limited aggregation(DLA), etc. The common feature shared by these 
processes is the existence of an evolving interface. 
The problem of the evolution of the interface in these processes attracts 
considerable attention (see, for example, the book~\cite{Pelce}) both because 
of its great practical importance and because of its connections with such fields as dynamical chaos, nonequilibrium physics, and fractal growth 
(see DLA~\cite{Witten}).

A general scheme for these processes is as follows. 
There is a linear PDE (frequently of the second order) for the scalar field 
determining the process. For example, this is the diffusion equation for the 
Stefan problem and the Laplace or Helmholtz equations for electrodeposition. 
This scalar field is temperature in the Stefan problem, pressure in flows 
through porous media, concentration of the nutrient for bacterial growth, 
electrostatic potential in 
electrodeposition, probability of the next jump in DLA, etc. Appropriate 
boundary conditions are imposed both on the moving part of the boundary (the interface) and non-moving part of the boundary (the outer walls). In addition, a law of interface motion is given in terms of the local behavior of the main scalar
field. (Typically the local velocity of the interface is 
proportional to the gradient of the scalar field near the interface). The main
question is: `what is the evolution of the interface?'

It is remarkable that some of the problems mentioned above are exactly 
solvable in two dimensions~\cite{ME}--\cite{David} 
despite the nonlinearity of these processes. These problems were solved with the help of time--dependent conformal mapping which cannot be extended to 3D 
except for a few trivial cases. In 2D, it was found that such 
processes as two--phase flows in porous media, electrodeposition, and slow 
solidification in a supercooled liquid or from a supersaturated binary 
solution are governed by an infinite number of the constants of motion which 
were obtained explicitly in several special cases~\cite{ME}--\cite{David}. 
These constants of motion are related to the conserved moments proposed 
by Richardson~\cite{Richardson}. This invariance is quite subtle and 
disappears when realistic physical perturbations such as surface tension or 
random noise are added. 

In the 3D case very few exact analytical results are known: a constant-velocity paraboloid~\cite{iv}, and a self-similarly growing ellipsoid~\cite{sam2}.
The only known way to obtain these solutions is by using one of the eleven 
coordinate system in which the 3D Laplace equation is separable~\cite{mors}. 
One then considers level surfaces as moving interfaces. It is clear that this method does not work when the shapes are time-dependent. The traditional attitude is that the main obstacle in obtaining nonperturbative exact results in 3D is the lack of (nontrivial) conformal mappings unlike in the 2D case. But is not this statement too strong?

This article is a natural extension of previous work~\cite{ME} to more 
general and realistic {\it multidimensional growth} problems; and not only for the Laplace equation as was done in~\cite{ME}--\cite{David}, but also for 
{\it arbitrary elliptic equations} describing for example, inhomogeneous and 
anisotropic diffusion in solidification, inhomogeneous dielectric functions 
and screening in electrodeposition, and inhomogeneous viscosity for flows 
through porous media. It turns out that these nonlinear processes also possess
remarkable properties (an infinite number of conservation laws) similar to the ones mentioned in \cite{ME}, \cite{Richardson} for the 2D Laplacian case; and these properties do not depend on the dimension of the process considered. Thus we show that, contrary to the traditional attitude, {\it we do not need a conformal mapping for this invariance. Rather this invariance originates from the more general property: the elliptic nature of the equation for the scalar field}.

Let us state now the following $D$--dimensional problem:
 
\begin{equation} 
L(u)={\rm div}(p({\bf r})\,{\rm grad}u({\bf r})) + q({\bf r})u({\bf r}) = 0
\end{equation} 
for ${\bf r} \in B\subset R^D$ where the domain $B$ is bounded by the 
nonmoving exterior boundary, $\Sigma$, and by 
the moving interior boundary $\Gamma (t)$ ($t$ is time) which is the 
interface separating the domains $B$ and $A$. 
An interior domain $A$ contains the origin and is surrounded by the moving interface $\Gamma(t)$ as shown at the Fig.1. 
Here $p$ and $q$ are given functions 
of ${\bf r} = (x_1,x_2,...,x_D)$.
The boundary conditions imposed on $u$ are:

\begin{equation}
\partial_n u|_\Sigma = G(\Sigma)\quad,
\end{equation}                                     

\begin{equation}
u(\Gamma(t))=0\quad.
\end{equation}                                                         
The left--hand side (LHS) of (2) means the normal component of grad($u$) 
evaluated at $\Sigma$.

There can also exist point--like sources and sinks in the domain, $B$, 
located at
${\bf r}_k$ and having strengths, $s_k$, ($k=1,2,...,N$), so that near 
${\bf r}_k \,, u({\bf r})$ diverges and satisfies 

\begin{equation}
u = s_k/{|{\bf r} - {\bf r}_k|^{D-2}} +{\rm~ smooth\,\, function}
\end{equation}                     
\noindent
if $D>2$, or

\centerline {$u = s_k\,$log$|{\bf r} - {\bf r}_k|  +  $ smooth function}  

\noindent                
if $D=2$.

The law of motion of $\Gamma(t)$ is

\begin{equation}
v_n = -p({\bf r}) \,\partial_n u|_{\Gamma(t)}\quad   , 
\end{equation}                           
\noindent
where the LHS is the normal component of the velocity of $\Gamma(t)$.

Equations (1) and (5) together with the boundary conditions (2)--(4) complete 
the mathematical description of the motion of $\Gamma(t)$. If, for example, 
$D=3,\, p({\bf r})=$const, and $q({\bf r})=0$, this describes slow 
solidification or two--phase flow in porous media.     

In this paper, I show that, in spite of the complexity and nonlinearity of the 
processes described by eqs.~(1)--(5), these processes are governed by an 
infinite number of conservation laws. Namely, if the outer
boundary, $\Sigma$, is very far from the origin there is an infinite set of 
numbers $C_l\,\, ( l = D-1,D,D+1,...,\infty )$ defined as

\begin{equation}
C_l \equiv dM_l/{dt} \equiv d/{dt} (\int_B\psi_l\,d^D{\bf r})  
\end{equation}

which are conserved during the evolution of the hypersurface\footnote{We do not consider here the passing of the interface through the singularities.}, $\Gamma(t)$, and equal

\begin{equation}
C_l = 2\frac{\pi^{D/2}}{\Gamma(D/2)}\sum\nolimits_{k=1}^N s_k p({\bf r}_k) 
\psi_l({\bf r}_k)\ \ .
\end{equation}                  

Here, $\Gamma(n)$ is the Gamma function ({\bf not} the interface $\Gamma(t)$) 
and $\psi_l$ is the arbitrary solution of $L(u)=0$ which decays at infinity no 
slower than $r^{-l}$ and having singularity only at the origin\footnote{The 
functions $\psi_l$ as well as quantities $C_l$ and $M_l$ are labeled in general by more than one number (which is $l$ here). See for example the Eq.(11) below. But for simplicity and without the loss of generality we drop all labels except the $l$ almost everywhere.}. If $\,q({\bf r})=0$ we have one more conserved 
quantity: $C_0$, which is the rate of the change of the volume of $B$ when $L$ 
is Laplacian, and which satisfies 

\begin{eqnarray}
C_0 & = & dM_0/{dt}  = d/{dt} (\int_B p({\bf r})\,d^D{\bf r}) \nonumber 
\\
& = & 2\frac{\pi^{D/2}}{\Gamma(D/2)}\sum\nolimits_{k=1}^N s_k p({\bf r}_k) + 
\int_\Sigma G(\Sigma)\,p(\Sigma)\,d\Sigma\quad,
\end{eqnarray}
\noindent
Here we took $\psi_0 = 1$ which is a solution of $L(u)=0$ when $q({\bf r})=0$. 

{\it We think that the knowledge of $C_l$'s together with the initial $M_l$'s defined in (6) (the latters are uniquely determined by the initial shape of the interface) could describe the whole moving shape in many of important physical
cases. We consider eqs.~(7) and (8) as the main result of this work}.

Formula (7) follows immediately from the following considerations:

\noindent
Since 

$$\int_{B(t+dt)} \psi_l\,d^D{\bf r} -\int_{B(t)} \psi_l\,d^D{\bf r} = 
\int_{\Gamma(t)} \psi_l\,v_n\,d\Gamma\,dt$$ 

\noindent
we have 

$$d/{dt} (\int_{B(t)} \psi_l\,d^D{\bf r}) =
\int_{\Gamma(t)} \psi_l\,v_n\,d\Gamma$$          
                                                              
\noindent
Further, because of (5), it equals

$$=\int_{\Gamma(t)} (-p\,\psi_l\,\partial_n\,u)d\Gamma$$

\noindent
and, finally, in view of (3) this expression is 

\begin{eqnarray}
& = &\int_{\Gamma(t)} p(u\,\partial_n\,\psi_l - \psi_l\,\partial_n\,u)\,
d\Gamma\nonumber \\
& = & \int_{\Gamma(t)} 
(\sqrt p\,u\,
\partial_n(\sqrt p\,\psi_l) - \sqrt p\,\psi_l\,\partial_n( \sqrt p\,u))\,
d\Gamma\quad.
\end{eqnarray}                                                     

Applying Green's theorem to the functions $\sqrt p\,u$ and 
$\sqrt p\,\psi_l$ we find that the RHS of eq.~(9) is given by 

\begin{eqnarray}
\int_B 
{\rm div}(\sqrt p\,u\,{\rm grad}(\sqrt p\,\psi_l) & - &\sqrt p\,\psi_l\,
{\rm grad}(\sqrt p\,u))\,d^D\,{\bf r}\nonumber \\
& + & \int_{\Sigma} p(u\,\partial_n\,\psi_l -\psi_l\,\partial_n\,u)\,
d\Sigma  \\ 
& + &\sum\nolimits_{k=1}^N \int_{\gamma_k} 
p(u\,\partial_n\,\psi_l - \psi_l\,\partial_n\,u)\, d\Sigma\quad.\nonumber
\end{eqnarray}              

\noindent
Here the summation is over the point--like charges, $s_k$, 
and $\gamma_k$ denotes the surface of the infinitesimal hypersphere around the 
${\bf r}_k$. 

Considering the RHS of (10) one can see that: 

(i) the  volume integral over $B$ vanishes because of (1);

(ii) the surface integral over $\Sigma$ also vanishes if the outer boundary, 
$\Sigma$, is far removed from the center and if $\psi_l$ decays at infinity 
stronger than $1/r^{D-2}$. (When $\psi_l = 1$, this integral is not zero but 
equals $\int_{\Sigma} G(\Sigma)\,p(\Sigma)\,d\Sigma$ as it is in the RHS of 
(8));

(iii) the contribution of the first integrand to the surface integral over 
the $\gamma_k$ is zero, while the integral from the second term equals 
\begin{displaymath}
- 2\frac{\pi^{D/2}}
{\Gamma(D/2)}s_k\,p({\bf r}_k)\,\psi_l({\bf r}_k) 
\end{displaymath}
since $p({\bf r})$  and $\psi_l({\bf r})$ are regular near the ${\bf r}_k$ 
and due to Gauss's theorem. 

Thus the RHS of (10) equals the RHS of (7) (or (8) when $q({\bf r}) = 0$ and 
$l=0$), so we have obtained the infinite set of the conserved quantities, $C_l$, if the sources and sinks are nonmoving (i.e. when $s_k$ and ${\bf r}_k$ are 
time--independent). Moreover, if $q({\bf r}) = 0$ and no point--like 
singularities are in the domain $B$ (i.e. if all $s_k = 0$), then 
{\it all the dynamics of the interface $\Gamma(t)$ has been reduced to the linear time--dependence of only one ``moment'', $M_0$, which is the volume of the 
phase $B$ if $L$ is Laplacian. All higher moments, $M_l$, are constant in time}. Note also that {\bf all moments $M_l$ have a purely geometrical nature}, and thus carry information about the moving shape.

It is worth mentioning the physical interpretation of the derived invariants, $M_l$, in the special case when the operator $L$ is Laplacian: $p = 1,\, q = 0$\footnote{ In this case it was noted by Howison~\cite{sam2} that the integral of a harmonic function over the infinite domain enclosed the moving bubble is a constant in time.}. For $D =2$, if the $\psi_l$ are chosen as $\psi_l = z^{-l}$ where $z=x+iy$, these integrals coincide with those previosly found in explicit form~\cite{ME} via the coefficients of the appropriate conformal map. These are analogs of the Richardson moments, whose invariance was found earlier for the interior Hele--Shaw problem ~\cite{Richardson}.

In the 3D Laplacian case, one can choose the $\psi_l$ to be a set of spherical functions:

\begin{equation}
\psi_l^{(m)} = P_{(l-1)}^{(m)}(\theta)\,e^{im\phi}/{r^l}  \quad. 
\end{equation}               

Here, $r,\,\theta$, and $\phi$ are the polar coordinates and 
$P_l^{(m)}(\theta)$ are the associated Legendre polynomials. In this case, the 
moments, $M_l^ {(m)}$, are the coefficients of the multipole expansion of the 
Newtonian potential at an arbitrary point of the empty interior domain $A$, if 
the potential is created by the mass uniformly occupying the domain $B$.
Thus the question of whether one can recover the moving shape using only the 
numbers, $M_l$, introduced in (7) 
is now reduced to the classical inverse potential problem~\cite{Novikoff} 
for the reconstruction of the shape of 
a body of constant density from its Newtonian potential. Our case corresponds to the exterior problem (where the potential is given in the empty hollow of the 
body: in the phase $A$).
The author believes that the connection 
between pattern formation studies and 
the inverse potential problem is especially important and deserves close 
attention. We do not discuss this problem here. Rather we merely note that in 3D (unlike the 
2D case) there is no description of a body (with the exception of the sphere) in terms of a finite number of nonzero moments, $M_l$.
(For a detailed description of these difficulties see 
\cite{Herglotz}\footnote{I am grateful to M.~Brodsky who brought my attention 
to this book.}).

Note that it is also possible in the general elliptic case (when $L$ is not 
Laplacian) to preserve the interpretation of the conserved quantities $M_l$ 
as coefficients of the orthogonal expansion by choosing the Green's function 
$G({\bf r},{\bf r}_1)$ of the operator, $L$, to be the integrand in (7), since 
$G({\bf r},{\bf r}_1)$ satisfies the conditions imposed on the $\psi_l$ if 
${\bf r} \in B$ and ${\bf r}_1 \in A$. Using the Green's function expansion 
and by choosing the orthogonal set of the eigenfunctions $\psi_l$ of ~$L(\psi_l)=0$ bounded at infinity and divergent at the origin and the complementary set of 
eigenfunctions $\tilde \psi_l$ of the same equation but with opposite 
asymptotics, we have

$$U({\bf r}) = 
\sum\nolimits_l \tilde \psi_l ({\bf r}) d/{dt} (\int_B \psi_l ({\bf r}_1)\,
d^D{\bf r}_1)  \quad.
$$                                                

We remark that our main result (7) holds also for 
time--dependent $s_k$ and ${\bf r}_k$ (since we never used the 
time--independence of $s_k$ and ${\bf r}_k$ in obtaining (7)). Although the 
$C_l$ are no longer conserved, the problem is still integrable, since $C_l$ 
are known functions of time (if the time--dependence of $s_k$ and ${\bf r}_k$ 
is given). Thus the moments, $M_l$, are easily controlled parameters, namely 
they are just primitives of the time--dependent RHS of (7). In this way, one 
might be able to govern the motion of the interface by the proper choice of 
the $s_k$ and ${\bf r}_k$.

It should also be mentioned that there could be a few exceptions among the 
$C_l$'s (only one in the 3D Laplacian case) for which eq.~(7) is not valid and 
thus they may not be conserved.
These nonconserved $C_l$'s correspond to the $\psi_l$'s decaying at infinity 
but not stronger than $r^{(2-D)}$.  For the 3D Laplacian case the only 
nonconserved quantity is $C_1$, which corresponds to $\psi_1 = r^{-1}$. Since 
we do not know the time--dependence of $C_1$ one could suppose that the 
description of the interface is now less complete. However, although $C_1$ 
is not conserved, we have the conserved quantity $C_0$ (see eq.~(8)). 
In other words we think that the set \{$M_1,\,M_2,\,M_3,\,...$\} describes 
the shape with the same degree of completeness as the set 
\{$M_0,\,M_2,\,M_3,\,...$\}. This question concerning nonconserved 
quantities  among the $C_l$'s does not arise for the interior problem when 
the scalar field $u$ is given in the phase enclosed by the moving interface $\Gamma(t)$ (in our case in $A$ instead of $B$). 

In conclusion we pose several questions that arise from these studies and which we beleive merit attention:

1) Does the relation expressed by (7) really mean complete integrability
 of the multidimensional growth? 

2) If yes, to which nonlinear evolutionary PDE's do these constants $C_l$ 
correspond?  

3) Is there a Hamiltonian structure for these systems?

4) Do finite--time singularities (cusps?) exist here, as in the 2D case?
 And, if yes, how may surface tension regularise them?

5) How can one recover the moving shape from the given set of moments, $M_l$?

\vskip .5truein
\noindent
{\bf Acknowledgments}

It is a pleasant obligation to thank W.~Bruno, D.~Leschiner, and J.~Pearson for 
a useful discussion, and I.M.~Gelfand for his interest in this activity and 
for his remarks. This work was begun at the Department of Mathematics at Rutgers University and was completed at the Center for Nonlinear Studies in Los Alamos. I wish to thank both of these institutions and J.~Lebowitz and G.~Doolen for their generous hospitality. 

This work was partially supported by AFOSR Grant 91--0010.


\begin{thebibliography}{Richardson}

\bibitem{Pelce} ``Dynamics of Curved Fronts'' 
ed. by P.~Pelce, Academic Press, Inc. (1988)
\bibitem{Witten}T.A.~Witten, L.M.~Sander,  Phys.~Rev.~Lett. {\bf 47}, 1400 (1981) 
\bibitem{ME}M.B.~Mineev, Physica {\bf D 43}, 288 (1990) 
\bibitem{Sam}S.D.~Howison, J.~Fluid Mech. {\bf 167}, 439 (1986)  
\bibitem{David}D.~Bensimon, P.~Pelce, Phys.~Rev. {\bf A 43}, 4477 (1986)
\bibitem{Richardson}S.~Richardson, J.~Fluid Mech. {\bf 56}, 609 (1972) 
\bibitem{iv}G.P.~Ivantsov, Doklady Acad.~Nauk SSSR, {\bf 58}, No.~4, 567 (1947)
\bibitem{sam2}S.D.~Howison, Proc.~Royal Society of Edinburgh, {\bf 102A}, 141 (1986) 
\bibitem{mors}P.M.~Morse and H.~Feshbach, ``Methods of Theoretical Physics'', McGraw Hill Book Co., Inc. (1953)
\bibitem{Novikoff}P.S.~Novikoff, Doklady AN SSSR {\bf 18}, 165 (1938) 
and V.~Strakhov, M.~Brodsky, SIAM J.~of Appl.~Math. {\bf 46}, n.2, 324 (1986)
\bibitem{Herglotz}G.~Herglotz, ``\"Uber die analytische Fortsetzung des 
Potentials ins Innere der anziehenden Massen'', Teubner--Verlag (1914)

\end{thebibliography}
\end{document}